\documentclass[preprint,aps,prd,superscriptaddress,nofootinbib,tightenlines]{revtex4}

\usepackage{amsfonts}
\usepackage{multirow}
\usepackage{mathrsfs}
\usepackage{graphicx}
\usepackage{amsmath}
\usepackage{booktabs}
\usepackage{amssymb}
\usepackage{bm}
\usepackage{bbm}
\usepackage{color}
\usepackage{hyperref}
\usepackage{ulem}
\usepackage{subfig}
\usepackage{caption}
\usepackage{slashed}
\usepackage{float}
\usepackage{makecell}

\def\OMIT#1{}

\newcommand{\beq}{\begin{equation}}
\newcommand{\eeq}{\end{equation}}
\newcommand{\bqa}{\begin{eqnarray}}
\newcommand{\eqa}{\end{eqnarray}}

\newcommand{\bseq}{\begin{subequations}}
\newcommand{\eseq}{\end{subequations}}

\newcommand{\dd}{{\mathrm{d}}}

\newcommand{\mel}[3]{\left\langle{#1}\left\lvert{#2}\right\rvert{#3}\right\rangle}
\newcommand{\expval}[1]{\left\langle{#1}\right\rangle}
\newcommand{\abs}[1]{\left\lvert{#1}\right\rvert}

\newcommand{\triplet}{\ensuremath{\mathbf{\bar{3}}\otimes\mathbf{3}}}
\newcommand{\sextet}{\ensuremath{\mathbf{6}\otimes\mathbf{\bar{6}}}}

\preprint{JLAB-THY-23-3965}

\begin{document}


\title{Photoproduction of fully charmed tetraquark at electron-ion colliders}

\author{Feng Feng~\footnote{F.Feng@outlook.com}}
\affiliation{China University of Mining and Technology, Beijing 100083, China\vspace{0.2 cm}}
\affiliation{Institute of High Energy Physics, Chinese Academy of
  Sciences, Beijing 100049, China\vspace{0.2 cm}}

\author{Yingsheng Huang~\footnote{yingsheng.huang@northwestern.edu}}
\affiliation{High Energy Physics Division, Argonne National Laboratory, Argonne, IL 60439, USA}
\affiliation{Department of Physics \& Astronomy,
  Northwestern University, Evanston, IL 60208, USA}

\author{Yu Jia~\footnote{jiay@ihep.ac.cn}}
\affiliation{Institute of High Energy Physics, Chinese Academy of
  Sciences, Beijing 100049, China\vspace{0.2 cm}}
\affiliation{School of Physical Sciences, University of Chinese Academy of Sciences,
  Beijing 100049, China\vspace{0.2 cm}}

\author{Wen-Long Sang~\footnote{wlsang@swu.edu.cn}}
\affiliation{School of Physical Science and Technology, Southwest University, Chongqing 400700, P.R. China}

\author{De-Shan Yang\footnote{yangds@ucas.ac.cn}}
\affiliation{School of Physical Sciences, University of Chinese Academy of Sciences,
  Beijing 100049, China\vspace{0.2 cm}}
\affiliation{Institute of High Energy Physics, Chinese Academy of
  Sciences, Beijing 100049, China\vspace{0.2 cm}}

\author{Jia-Yue Zhang~\footnote{jzhang@jlab.org}}

\affiliation{Theory Center, Jefferson Lab, Newport News, Virginia 23606, USA}
\affiliation{Institute of High Energy Physics, Chinese Academy of Sciences, Beijing 100049, China\vspace{0.2 cm}}

\date{\today}

\begin{abstract}
In this work, we investigate the inclusive photoproduction of the $C$-odd, $S$-wave fully charmed tetraquark at electron-ion colliders
within the nonrelativistic QCD (NRQCD) factorization framework, at the lowest order in velocity and $\alpha_s$.
The value of the NRQCD long-distance matrix element is estimated from two phenomenological potential models.
Our  studies reveal that the photoproduction of the $1^{+-}$ fully charmed tetraquark may be difficult to observe at \texttt{HERA} and the \texttt{EicC};
nevertheless, its observation prospect at the \texttt{EIC} appears to be bright.
\end{abstract}

\maketitle
\newpage

\section{Introduction}

Quarkonium photoproduction at the electron-proton collider has been an interesting topic, which provides an ideal platform to test the
quarkonium production mechanism and extract the gluon content of the proton~\cite{Brambilla:2010cs,Brambilla:2014jmp}.
With an almost on-shell photon emitted from the incident electron beam, the $ep$ collider provides a much cleaner environment
to study the quarkonium production than the hadron colliders. During the past few decades
a number of experimental and theoretical efforts have been devoted to studying the $J/\psi$ photoproduction at
\texttt{HERA}~\cite{hep-ex/9603005,hep-ex/9708010,hep-ex/0211011,hep-ex/0205064,0906.1424,1002.0234,1211.6946,hep-ph/9511433,hep-ph/9508409,hep-ph/9601276,hep-ph/9602223,hep-ph/0512194,hep-ph/0602091,0901.4352,0901.4749,0909.2798,1009.5662,1105.0820,1903.09185,Flore:2020jau,2005.10832,2112.05060}.
Moreover, studies of quarkonium photoproduction are also of high priority in the upcoming next-generation electron-ion collider programs,
exemplified by the US Electron-Ion Collider (\texttt{EIC})~\cite{1212.1701, 2103.05419} and
the Electron-ion collider in China (\texttt{EicC})~\cite{2102.09222}. For example, the near-threshold $J/\psi$ photoproduction at the \texttt{EIC} and \texttt{EicC}
has been advocated as the gold-plated process to infer the QCD trace anomaly contribution to the nucleon mass and
extract the nucleon's generalized parton distribution
functions~\cite{1212.1701, 2103.05419,2102.09222,nucl-th/9601029,hep-ph/9901375,Sun:2021pyw,2103.11506,2305.06992}.

In addition to helping us study the quarkonium production, the electron-ion colliders also serve a fruitful platform to study
the production of exotic hadrons such as charmonium-like $XYZ$ exotic states~\cite{1212.1701, 2103.05419}.
The goal of this work is to study the inclusive photoproduction of a special class of exotic hadrons, {\it i.e.}, the fully charmed tetraquark (dubbed $T_{4c}$ henceforth) and assess 
its observation prospects at various electron-ion colliders.

An unexpected discovery of the $X(6900)$ resonance in the di-$J/\psi$ invariant mass spectrum by the \texttt{LHCb} collaboration in 2020~\cite{Aaij:2020fnh}, later confirmed by
both the \texttt{ATLAS} and \texttt{CMS} Collaborations~\cite{ATLAS:2023bft,CMS:2023owd}, has triggered a flurry of intensive theoretical investigation on the properties of
the close relatives of quarkonium which are composed of four heavy quarks.
Although there are several alternative interpretations such as charmonia molecules or hybrids,
it is most natural to regard the $X(6900)$ resonance as a strong candidate for the compact $T_{4c}$ state.
As a matter of fact, investigations of the fully heavy tetraquark states date back to the 1970s~\cite{Iwasaki:1976cn,Chao:1980dv,Ader:1981db},  long before the discovery of the $X(6900)$.
The predictions of mass spectra and decay properties of fully heavy tetraquarks have been pursued through various phenomenological models, including quark potential models \cite{Becchi:2020uvq, Lu:2020cns, liu:2020eha, Karliner:2020dta, Zhao:2020nwy, Zhao:2020cfi, Giron:2020wpx, Ke:2021iyh, Gordillo:2020sgc, Jin:2020jfc, Mutuk:2022nkw, Wang:2022yes}, QCD sum rules \cite{Chen:2020xwe, Wang:2020ols, Yang:2020wkh, Wan:2020fsk, Zhang:2020xtb} and effective field theories~\cite{Sang:2023ncm,2311.01498}.
In contrast, the investigation on the production mechanism of fully heavy tetraquarks is mainly based on color evaporation models and duality relations \cite{Karliner:2016zzc, Berezhnoy:2011xy, Berezhnoy:2011xn, Becchi:2020mjz, Becchi:2020uvq, Maciula:2020wri, Carvalho:2015nqf, Goncalves:2021ytq}.

From the theoretical perspective, fully heavy tetraquarks are among the simplest exotic hadrons to analyze. Due to the heavy quark mass being much greater than $\Lambda_{\rm QCD}$,
the $T_{4c}$ may be viewed as a composite system made of four nonrelativistic charm and anticharm quarks.
The leading Fock state of the $T_{4c}$ is simply $\vert cc\bar{c}\bar{c}\rangle$, without the contamination from the light quarks and gluons.
This is quite analogous to the ordinary charmonia, whose leading Fock component is simply $\vert c\bar{c}\rangle $.
The similarity between the fully heavy tetraquarks and heavy quarkonia strongly indicates that
the theoretical tools developed in the past to tackle quarkonium may also be transplanted to
describe the fully heavy tetraquarks.

Recently, several groups have attempted to investigate the $T_{4c}$ production in the spirit of the nonrelativistic QCD (NRQCD)
factorization approach~\cite{Ma:2020kwb, Feng:2020riv, Feng:2020qee, Huang:2021vtb, Zhu:2020xni,Feng:2023agq}.
As an effective-field-theory-based modern method,
NRQCD factorization has been extensively employed to describe various quarkonium production and decay processes~\cite{hep-ph/9407339}.
Ma and Zhang studied the inclusive production of $T_{4c}$ at the \texttt{LHC} and conducted a numerical study of the dependence of the
ratio $\sigma(2^{++})/\sigma(0^{++})$ on $p_T$~\cite{Ma:2020kwb}.
Zhu considered the $gg\to T_{4c}$ channel and predicted the low-$p_T$ spectrum of the $T_{4c}$ at the \texttt{LHC},
taking small-$p_T$ resummation into account~\cite{Zhu:2020xni}.
Feng {\it et al.} explicitly constructed the NRQCD operators relevant to the $S$-wave $T_{4c}$ production,
and established the connection between the long-distance matrix elements (LDMEs) and the tetraquark wave functions at the origin~\cite{Feng:2020riv}.
The authors predicted the $T_{4c}$ hadroproduction rates via both the fragmentation mechanism~\cite{Feng:2020riv} and fixed-order NRQCD calculation~\cite{Feng:2023agq},
as well as the exclusive radiative production and inclusive production of $T_{4c}$ at $B$ factories~\cite{Feng:2020qee, Huang:2021vtb}.
They demonstrated that, compared with the $e^+e^-$ collider, the hadron colliders have much brighter potential for observing
the fully charmed tetraquarks.
The goal of this work is to further investigate the photoproduction of the $T_{4c}$ at electron-proton colliders
such as \texttt{HERA}, \texttt{EIC}, and \texttt{EicC}.
In particular, we are interested in the large-$p_T$ regime where NRQCD factorization can be safely applied and
the resolved-photon contribution gets heavily suppressed.

The rest of this paper is organized as follows.
In Sec.~\ref{sec:factorization},
with the aid of the equivalent photon approximation (EPA), we express the inclusive $T_{4c}$ production at the electron-ion colliders
in NRQCD factorization~\cite{Feng:2023agq}.
In Sec.~\ref{sec:SDC}, we compute the NRQCD short-distance coefficient (SDC) at leading order (LO) in $\alpha_s$ and $v$ via perturbative matching procedure.
In Sec.~\ref{sec:phenomenology}, we estimate the NRQCD long-distance matrix element (LDME) through two phenomenological potential models.
We then make concrete predictions about the $p_T$ distributions, integrated cross sections, and event yields
for $T_{4c}$ photoproduction at \texttt{HERA}, \texttt{EIC}, and \texttt{EicC}.
Finally, we summarize our findings in Sec.~\ref{sec:summary}.

\section{NRQCD Factorization for photoproduction of $T_{4c}$  \label{sec:factorization}}

The photoproduction process at electron-ion colliders can be well approximated by the EPA, also known as the Weizs\"acker-Williams approximation~\cite{vonWeizsacker:1934nji, Williams:1934ad, Budnev:1975poe}.
In this approximation, the low-virtuality photon entering the hard-scattering process is treated as a quasireal particle,
thus reducing the $2\to 3$ process to a $2\to 2$ one with the on-shell photon in the initial state.
Such an approximation has been routinely used to predict the inclusive photoproduction rate of an identified hadron,
such as the large-$p_T$ photoproduction of $J/\psi$~\cite{Flore:2020jau} and $D^{*\pm}$~\cite{Kniehl:1996we}.

We use the symbol $\sqrt{s}$ to signify the center-of-mass energy of the $ep$, and we use the symbol $W$ to signify the center-of-mass energy of
the $\gamma p$ subsystem. $x_\gamma$ is defined to be the momentum fraction carried by the photon relative to the incident electron, similar to the momentum fraction of a parton inside the proton.  
As such, we have $W= \sqrt{x_\gamma s}$.
It is also convenient to define the elasticity parameter $z\equiv P_{T_{4c}} \cdot P_p /P_\gamma\cdot P_p$, which can be interpreted as the
fraction of the photon energy taken up by the tetraquark in the proton rest frame. 
As $z$ approaches unity, one usually is concerned with the contamination from the diffractive contributions.
To be cautious, one may also need to include the resolved photon contribution, where the photon also entails nontrivial partonic distributions.
Nevertheless, the focus of this work is on the $T_{4c}$ production in the large-$p_T$ region, where the diffractive and resolved photon contributions can be
largely removed by imposing the cuts on the $z$ parameter.

Utilizing the EPA~\cite{vonWeizsacker:1934nji, Williams:1934ad, Budnev:1975poe},
one can express the inclusive production rate of the $T_{4c}$ at the $ep$ collision
as~\cite{Flore:2020jau}
\beq
{d \sigma \over d z d p_T}  = \sum_i \int_{x_\gamma^{\min }}^1 d x_\gamma
{2 x_i p_T \over z(1-z)}\, f_{\gamma / e}(x_\gamma)
   f_{i / p}(x_i, \mu)
  \ \frac{\dd \hat{\sigma}(\gamma+i \rightarrow T_{4c}+j;\mu)}{\dd \hat{t}},
\label{EPA:QCD:factorization:formula}
\eeq
where $i,j=g, q$ represent the partons in QCD,  and $p_T$ denotes the transverse momentum of the $T_{4c}$.
$f_{\gamma/e}$ represents the electron's parton distribution function (PDF) to find a photon with definite momentum fraction, and $f_{i/p}$
denotes the standard proton PDF for finding a parton $i$ with a certain momentum fraction.
$\hat{\sigma}$ refers to the partonic cross section, with $\hat{s}$  and $\hat{t}$ denoting the partonic Mandelstam variables.
$\mu$ represents the QCD factorization scale.
The momentum fraction of the parton $i$ inside the proton is a function of $x_\gamma$, $z$, $s$, and $M_{T_{4c}}$,
$x_i= {M_T^2- z M_{T_{4c}}^2\over x_\gamma z(1-z)\,s}$, where the transverse mass $M_T \equiv \sqrt{M_{T_{4c}}^2 +p_T^2}$.
The minimal momentum fraction of the photon is given by $x_{\gamma}^{\textrm{min}}=\frac{M_T^2-M_{T_{4c}}^2 z}{s\,z(1-z)}$.
The photon flux $f_{\gamma/e}$ is determined by the EPA~\cite{Kniehl:1996we,Flore:2020jau}:
\begin{align}
  & f_{\gamma / e}\left(x_\gamma \right)=\frac{\alpha}{2 \pi} {\left[\frac{1+\left(1-x_\gamma\right)^2}{x_\gamma} \ln \frac{Q_{\max }^2}{Q_{\min }^2\left(x_\gamma\right)}+2 m_e^2 x_\gamma\left(\frac{1}{Q_{\max }^2}-\frac{1}{Q_{\min }^2\left(x_\gamma\right)}\right)\right]},
\end{align}
where $Q_{\textrm{min}}^2 (x_\gamma)=m_e^2x_\gamma^2/(1-x_\gamma)$ and $m_e$ is the electron mass.
The value of $Q_{\max }^2$ varies with experiments, with a typical magnitude of around a few $\mathrm{GeV}^2$. 

Since the $T_{4c}$ photoproduction induced by the light quark is suppressed by extra powers of $\alpha_s$,
in this work we only consider the dominant partonic channel $\gamma+g\to T_{4c}+g$.
Due to $C$-parity conservation, such a partonic process can only produce a vector $S$-wave tetraquark state $1^{+-}$ (denoted as $T_{4c}^{(1)}$ below)
at the lowest order in $\alpha_s$.\footnote{Note that the $S$-wave $T_{4c}$ family also includes the $0^{++}$ and $2^{++}$ states,
which can be produced at the \texttt{LHC} via the
partonic channel $gg\to T_{4c}+g$, but are not permissible in the photoproduction channel $\gamma g\to T_{4c}+g$.}

The partonic cross section ${\dd \hat{\sigma}}/{\dd \hat{t}}$ in \eqref{EPA:QCD:factorization:formula} still encapsulates
nonperturbative effects related to the hadronization into the tetraquark.
Owing to the asymptotic freedom of QCD, ${\dd \hat{\sigma}}/{\dd \hat{t}}$ can be further factorized into the product of the
perturbatively calculable SDC and the NRQCD LDME.
The former encodes the creation of four heavy quarks above the scale of $\Lambda_{\textrm{QCD}}$,
while the latter entails the formation of the tetraquark at the length scale of $1/\Lambda_{\textrm{QCD}}$.
According to the NRQCD factorization, the partonic cross section in \eqref{EPA:QCD:factorization:formula}
at LO in the velocity expansion can be expressed as~\cite{Feng:2020riv,Feng:2020qee,Huang:2021vtb,2304.11142}:
\begin{align}
  \dfrac{\dd\hat{\sigma}(\gamma g\to T_{4c}^{(1)}+X)}{\dd\hat{t}} =\frac{2M_{T_{4c}}}{m_c^{14}}F_{3,3}^{(1)}(\hat{s},\hat{t})
  \expval{{O}^{(1)}_{3,3}},
  \label{eq:factorized:cross:section}
\end{align}
where ${O}^{(1)}_{3,3}$ denotes the NRQCD production operator associated with the $\triplet$ color channel in the diquark-antidiquark basis,
and $F_{3,3}^{(1)}$ is the respective SDC function.
As we have restricted ourselves to focus only on the $S$-wave $1^{+-}$ tetraquark, the $\sextet$ channel does not contribute due to Fermi statistics.
The NRQCD production operator ${O}^{(1)}_{3,3}$ can be explicitly written as~\cite{Feng:2020riv,Huang:2021vtb}:
\begin{align}
& {O}^{(1)}_{3,3} = \mathcal{O}^{i;(1)}_{\triplet}\sum_X|T_{4c}^{(1)}+X\rangle\langle T_{4c}^{(1)}+X|\mathcal{O}^{i;(1)\dagger}_{\triplet},
\end{align}
with
\begin{align}
& \mathcal{O}^{i;(1)}_{\triplet}= -{\frac{i}{\sqrt{2}}} \left[\psi_a^T (i \sigma^2)\sigma^j\psi_b\right]\left[\chi_c^\dagger\sigma^k (i \sigma^2)\chi_d^*\right]\,\epsilon^{ijk}\;{\mathcal C}^{ab;cd}_{\triplet}.\label{NRQCD:composite:operators}
\end{align}
The color indices $a,b,c,d$ in Eq.~\eqref{NRQCD:composite:operators} run from $1$ to $3$, and the Cartesian indices $i,j,k$
run from $1$ to $3$. The rank-$4$ color tensor $\mathcal{C}$ is defined as
  \begin{align}
     & {\mathcal{C}^{ab;cd}_{\mathbf{3} \otimes\bar{\mathbf{3}}}\equiv \frac{1}{2} \epsilon^{abm}\epsilon^{cdn}\frac{\delta^{mn}}{\sqrt{N_c}}=\frac{1}{2\sqrt{3}}(\delta^{ac}\delta^{bd}-\delta^{ad}\delta^{bc})}.\label{color:tensor}
  \end{align}

\section{Calculation of NRQCD short-distance coefficient \label{sec:SDC}}

\begin{figure}[htpb]
\centering
\includegraphics[scale=0.7]{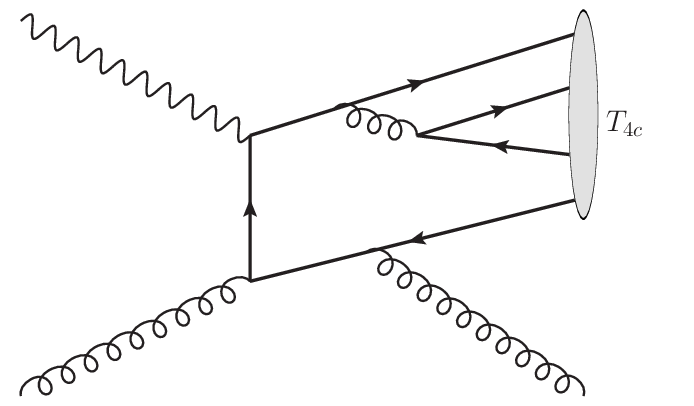}
\caption{A typical Feynman diagram for $\gamma g \to T^{(1)}_{4c}+g$ at lowest order in $\alpha_s$.
The blob represents the fully charmed tetraquark.}
\label{fig:feynmandiagram}
\end{figure}

Next we proceed to deduce the SDC $F_{3,3}^{(1)}$ in \eqref{eq:factorized:cross:section} through the tree-level matching procedure.
Since the SDC is insensitive to the long-distance dynamics,  one can replace a physical $T^{(1)}_{4c}$ with a free four-quark state carrying the quantum number $1^{+-}$, dubbed ${\cal T}_{4c}^{(1)}$.
By computing both sides of \eqref{eq:factorized:cross:section} in perturbative QCD and perturbative NRQCD, respectively,
one can readily solve the SDC.

The calculation of the perturbative NRQCD side is straightforward.
We normalize the fictitious ${\mathcal T}_{4c}^{(1)}$ state such that the vacuum-to-${\mathcal T}_{4c}$ matrix element reads
\begin{align}
  \mel{\mathcal{T}^{(1)}_{\triplet}}{\bm{\varepsilon}\cdot\mathcal{O}_{\triplet}^{(1)}}{0}= 4,
\end{align}
where $\bm{\varepsilon}$ represents the polarization vector of the fictitious tetraquark.

In the perturbative QCD side, there are more than 300 tree-level Feynman diagrams for $\gamma g\to {\cal T}_{4c}^{(1)}+g$, one of which is displayed in Fig.~\ref{fig:feynmandiagram}.
Since we are interested in the lowest order in $v$, we assume that all four $c$ quarks inside ${\cal T}_{4c}^{(1)}$ carry equal momentum and $M_{T_{4c}}\approx 4m_c$.
We employ the covariant projection technique to facilitate the calculation~\cite{Feng:2020qee,Feng:2023agq}.
Feynman diagrams and amplitudes are generated by the package \texttt{FeynArts}~\cite{Hahn:2000kx}, the Lorentz contraction and trace algebra are handled by
\texttt{FeynCalc}~\cite{Hahn:2000kx,Shtabovenko:2016sxi} and \texttt{HepLib}~\cite{Feng:2021kha}.
Upon squaring the amplitude,
we only sum over two transverse polarizations for the photon and gluons.
The gauge invariance has been verified by showing that the final expression of the unpolarized squared amplitude
does not depend on the arbitrary auxiliary four-vector introduced in the polarization sum formula.

We end up with the following expression of the SDC:
\begingroup
\allowdisplaybreaks
\begin{align*}
&F_{3,3}^{(1)}(\hat{s},\hat{t}) =
 \pi^3 e_c^2
  \alpha_s^4 r_s^2\left[72 r_t^8\left(5445-5298 r_t+1462 r_t^2-184 r_t^3+49 r_t^4\right)-432 r_t^7\left(-5445\right.\right. \\
& \left.+9879 r_t-5524 r_t^2+1030 r_t^3-112 r_t^4+42 r_t^5\right) r_s+2 r_t^6\left(3332340-8427078 r_t+8454303 r_t^2\right. \\
& \left.-4101132 r_t^3+1115650 r_t^4-253810 r_t^5+43627 r_t^6\right) r_s^2+2 r_t^5\left(5880600-17892198 r_t\right. \\
& \left.+25180533 r_t^2-22035111 r_t^3+12807704 r_t^4-4945126 r_t^5+1195291 r_t^6-138533 r_t^7\right) r_s^3 \\
& +r_t^4\left(14113440-49523400 r_t+83600442 r_t^2-101112318 r_t^3+94409051 r_t^4\right. \\
& \left.-60657225 r_t^5+24055510 r_t^6-5305354 r_t^7+505879 r_t^8\right) r_s^4+r_t^3\left(11761200-49523400 r_t\right. \\
& +95733756 r_t^2-135804348 r_t^3+164472260 r_t^4-151209848 r_t^5+91395217 r_t^6 \\
& \left.-33278237 r_t^7+6611864 r_t^8-555538 r_t^9\right) r_s^5+r_t^2\left(6664680-35784396 r_t+83600442 r_t^2\right. \\
& -135804348 r_t^3+186897370 r_t^4-206629419 r_t^5+164091573 r_t^6-86266517 r_t^7 \\
& \left.+27956171 r_t^8-5016861 r_t^9+381715 r_t^{10}\right) r_s^6+r_t\left(2352240-16854156 r_t+50361066 r_t^2\right. \\
& -101112318 r_t^3+164472260 r_t^4-206629419 r_t^5+187216756 r_t^6-119518674 r_t^7 \\
& \left.+52323094 r_t^8-14762980 r_t^9+2381419 r_t^{10}-165406 r_t^{11}\right) r_s^7+\left(392040-4267728 r_t\right. \\
& +16908606 r_t^2-44070222 r_t^3+94409051 r_t^4-151209848 r_t^5+164091573 r_t^6 \\
& \left.-119518674 r_t^7+59925804 r_t^8-20969265 r_t^9+4946107 r_t^{10}-698919 r_t^{11}+43850 r_t^{12}\right) r_s^8 \\
& +\left(-381456+2386368 r_t-8202264 r_t^2+25615408 r_t^3-60657225 r_t^4+91395217 r_t^5\right. \\
& -86266517 r_t^6+52323094 r_t^7-20969265 r_t^8+5682942 r_t^9-1042547 r_t^{10}+119941 r_t^{11} \\
& \left.-6480 r_t^{12}\right) r_s^9+\left(105264-444960 r_t+2231300 r_t^2-9890252 r_t^3+24055510 r_t^4\right. \\
& -33278237 r_t^5+27956171 r_t^6-14762980 r_t^7+4946107 r_t^8-1042547 r_t^9+135646 r_t^{10} \\
& \left.-10512 r_t^{11}+408 r_t^{12}\right) r_s^{10}+\left(-13248+48384 r_t-507620 r_t^2+2390582 r_t^3-5305354 r_t^4\right. \\
& \left.+6611864 r_t^5-5016861 r_t^6+2381419 r_t^7-698919 r_t^8+119941 r_t^9-10512 r_t^{10}+324 r_t^{11}\right) r_s^{11} \\
& \left.+\left(2-3 r_t+r_t^2\right)^2\left(882-1890 r_t+13277 r_t^2-21970 r_t^3+14354 r_t^4-4032 r_t^5+408 r_t^6\right) r_s^{12}\right] \\
& \times\left\{
  {1327104}
  \left(3-r_s\right)^2\left(2-r_s\right)^2\left(1-r_s\right)^2\left(r_s\left(2-r_t\right)-2 r_t\right)^2\left(3-r_t\right)^2\left(2-r_t\right)^2\left(1-r_t\right)^2\right.
\\
& \left.\times\left(r_s+r_t\right)^2\left(r_s\left(3-2 r_t\right)-3 r_t\right)^2\right\}^{-1},\stepcounter{equation}\tag{\theequation}
\label{eq:SDC:large:pT}
\end{align*}
\endgroup
where $e_c=\frac23 e$, $r_s={16m_c^2}/{\hat{s}}$ and $r_t={16m_c^2}/{\hat{t}}$.
Although the full expression is  somewhat lengthy, its asymptotic form of the SDC in large $p_T$ is exceedingly simple:
\begin{align}
F_{3,3}^{(1)}(\hat{s},\hat{t}) = \frac{
  {605}
   \pi ^3 \alpha_s^4 e_c^2
   m_c^8 \left(\hat{s}^2+\hat{s} \hat{t}+ \hat{t}^2\right)^2}
  {
    {1458}
      \hat{s}^4 \hat{t}^2 (\hat{s}+\hat{t})^2}+\mathcal{O}\left({\frac{m_c^9}{p_T^9}}\right).
\end{align}
In the exceedingly large-$p_T$ limit, this $1/p_T^8$ falloff is much more suppressed with respect to the $1/p_T^4$ scaling from the fragmentation mechanism.
Since we are working with the LO accuracy in $\alpha_s$, we are unable to incorporate the fragmentation contribution.
However, it is the moderate $p_T$ regime where the bulk of cross section lies and fragmentation approximation ceases to be applicable,
our fixed-order NRQCD prediction is expected to yield a reliable order-of-magnitude estimate.

\section{Phenomenology \label{sec:phenomenology}}

\begin{figure}[H]
  \subfloat[$\sqrt{s}=45\ \mathrm{GeV}$]{
 \includegraphics[width=0.49\linewidth]{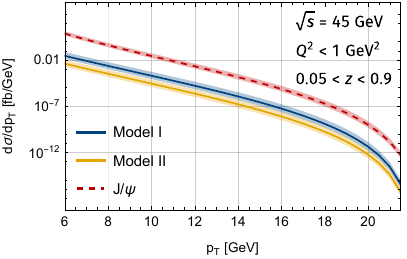}}\hfill
  \subfloat[$\sqrt{s}=63\ \mathrm{GeV}$]{
    \includegraphics[width=0.49\linewidth]{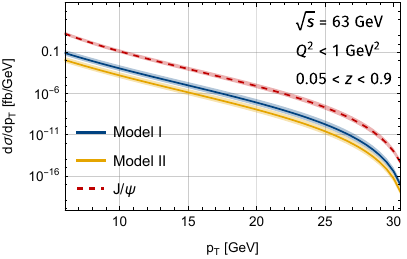}}\\
  \subfloat[$\sqrt{s}=105\ \mathrm{GeV}$]{
    \includegraphics[width=0.49\linewidth]{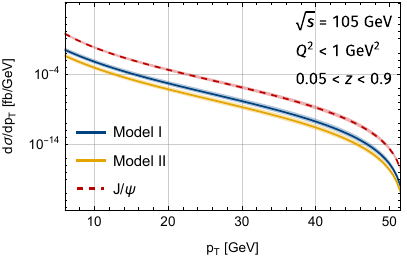}}\hfill
  \subfloat[$\sqrt{s}=140\ \mathrm{GeV}$]{
    \includegraphics[width=0.49\linewidth]{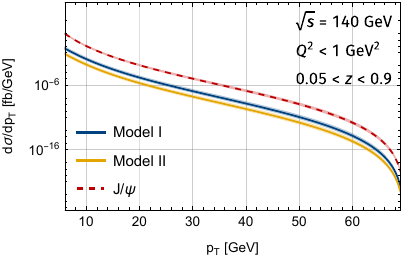}}
  \caption{Comparison of $p_T$ distributions of the vector $T_{4c}$ with LDMEs estimated from two phenomenological models, as well as with several beam-energy configurations of the proposed \texttt{EIC} as detailed in Table~\ref{tab:total:cross:section}. We also show the LO $p_T$ distribution of $J/\psi$ for comparison.
  \label{fig:model:compare:EIC}}
\end{figure}

\begin{figure}[H]
  \centering
  \subfloat[HERA: $\sqrt{s}=319\ \mathrm{GeV}$]{\includegraphics[width=0.49\linewidth]{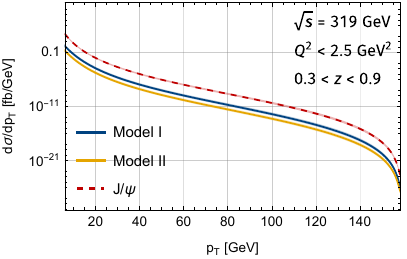}}\hfill\subfloat[EicC: $\sqrt{s}=20\ \mathrm{GeV}$]{
  \includegraphics[width=0.49\linewidth]{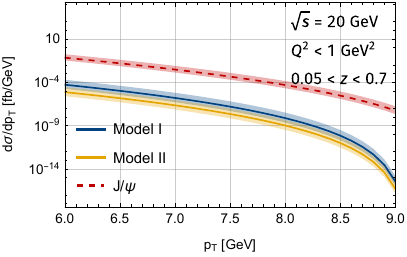}}
  \caption{Comparison of $p_T$ distributions of $T_{4c}$ with LDMEs estimated from two phenomenological models at \texttt{HERA} and the future \texttt{EicC}.
  \label{fig:model:compare:HERA}}
\end{figure}

\begin{figure}[H]
  \subfloat[$\sqrt{s}=45\ \mathrm{GeV}$]{
 \includegraphics[width=0.49\linewidth]{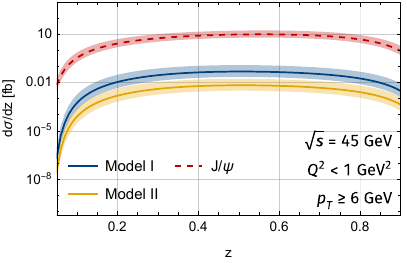}}\hfill
  \subfloat[$\sqrt{s}=63\ \mathrm{GeV}$]{
    \includegraphics[width=0.49\linewidth]{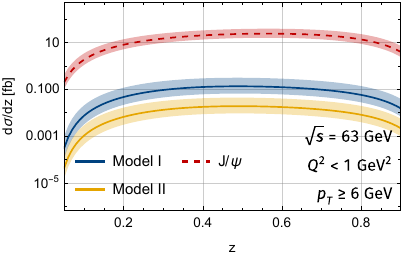}}\\
  \subfloat[$\sqrt{s}=105\ \mathrm{GeV}$]{
    \includegraphics[width=0.49\linewidth]{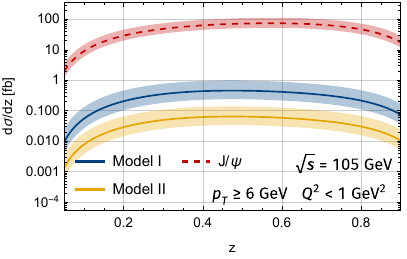}}\hfill
  \subfloat[$\sqrt{s}=140\ \mathrm{GeV}$]{
    \includegraphics[width=0.49\linewidth]{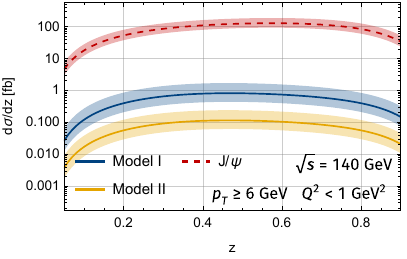}}
  \caption{Comparison of $z$ distributions of the vector $T_{4c}$ with LDMEs estimated from two phenomenological models, as well as with several beam-energy configurations of the proposed \texttt{EIC} as detailed in Table~\ref{tab:total:cross:section}.
  \label{fig:z:compare:EIC}}
\end{figure}

\begin{figure}[H]
  \centering
  \subfloat[HERA: $\sqrt{s}=319\ \mathrm{GeV}$]{\includegraphics[width=0.49\linewidth]{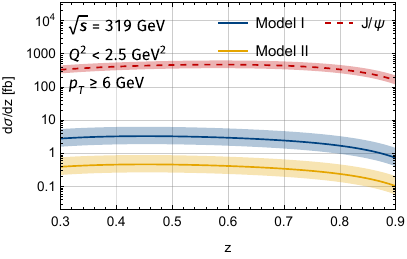}}\hfill\subfloat[EicC: $\sqrt{s}=20\ \mathrm{GeV}$]{
  \includegraphics[width=0.49\linewidth]{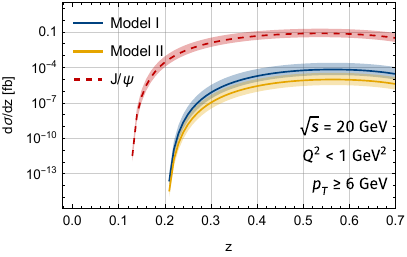}}
  \caption{Comparison of $z$ distributions of $T_{4c}$ with LDMEs estimated from two phenomenological models at \texttt{HERA} and the future \texttt{EicC}.
  \label{fig:z:compare:HERA}}
\end{figure}

In order to make concrete predictions for  photoproduction rates of the vector $T_{4c}$, we still need to know the concrete value of the LDME
that appears in the factorization formula (\ref{eq:factorized:cross:section}).
Since the LDME is a genuinely nonperturbative object, its value has to be ascertained through nonperturbative means.
The most reliable first-principle method to infer this matrix element is through the lattice NRQCD simulation.
Unfortunately, such a lattice study is absent thus far. As a temporary workaround, we appeal to the phenomenological potential models to estimate this LDME.
After applying the vacuum saturation approximation, we express the LDME in terms of the $T_{4c}$ wave function at the origin~\cite{Feng:2020riv,Feng:2020qee,Huang:2021vtb}:
\beq
\expval{{O}_{3,3}^{(1)}}\approx 48\,\abs{\psi_{\triplet}(\mathbf{0})}^2,
\label{LDME2wf}
\eeq
where $\psi(\mathbf{0})$ denotes the value of the Schr\"odinger wave function where all the $c$ quarks coincide in the same position.

In our numerical study, we resort to two types of potential models, referred to as model I~\cite{Lu:2020cns} and model II~\cite{liu:2020eha}.
With the aid of \eqref{LDME2wf}, these two models yield the following estimates for the LDME:
\beq
\textrm{Model I}: \expval{{O}_{3,3}^{(1)}}=0.078\;{\rm GeV}^9,\qquad\qquad \textrm{Model II}: \expval{{O}_{3,3}^{(1)}}=0.011\;{\rm GeV}^9.
\eeq

We then employ (\ref{eq:factorized:cross:section}) to calculate the $p_T$ spectrum of the vector $T_{4c}$ in the $ep$ collision,
at several electron-proton beam energy values, as indicated in Table~\ref{tab:total:cross:section}.
We take the charm quark mass $m_c=1.5\ \mathrm{GeV}$, $\alpha_s(M_Z)=0.1180$, and the default value of the renormalization and factorization scales $\mu=M_T$.
A rapidity cut $\abs{y}\leq 5$ is also imposed.
We utilize the \texttt{CT14lo} PDF set for the proton PDF~\cite{Dulat:2015mca}.
To account for the uncertainties arising from the higher-order QCD corrections, we slide the renormalization and factorization scales within the range $M_T/2 < \mu < 2 M_T$.

In Figs.~\ref{fig:model:compare:EIC} and \ref{fig:model:compare:HERA}, we present the $p_T$ distributions of the vector $T_{4c}$ 
at the \texttt{EIC}, \texttt{HERA} and \texttt{EicC}. We have considered four energy configurations of \texttt{EIC}, as detailed in Table~\ref{tab:total:cross:section}. 
To guarantee the events of the photoproduction type, we impose the cuts on the photon virtuality analogous to the case of the photoproduction of charmonia~\cite{Flore:2020jau}: 
$Q^2_{\mathrm{max}}= 2.5\ \mathrm{GeV}^2$ for \texttt{HERA} and $Q^2_{\mathrm{max}}=1\ \mathrm{GeV}^2$ for the \texttt{EIC} and \texttt{EicC}. 
 We also impose the cuts on the elasticity parameter to eliminate both the diffractive and resolved-photon contributions: $0.3<z< 0.9$ for \texttt{HERA}, 
 $0.05<z< 0.9$ for the \texttt{EIC}, and $0.05<z< 0.7$ for the \texttt{EicC}. {We also compare the cross sections of $T_{4c}$ with $J/\psi$. The photoproduction of $J/\psi$ at $ep$ colliders has been computed up to NLO in $\alpha_s$~\cite{0909.2798}. For simplicity, we only include $J/\psi$ production at LO in NRQCD, which is approximately 3 orders of magnitude larger than those of $T_{4c}$.}

{In Figs.~\ref{fig:z:compare:EIC} and \ref{fig:z:compare:HERA}, we present the $z$ distributions of the vector $T_{4c}$ 
at the \texttt{EIC}, \texttt{HERA} and \texttt{EicC}. The $p_T$ ranges are set to be larger than $6\ \mathrm{GeV}$. }

\begingroup
\setlength{\tabcolsep}{5pt} 
\begin{table}[H]
    \centering
    \begin{tabular}{cccccccccccc}
      \hline \hline
      &\multirow[c]{2}{*}{$\sqrt{s}$ [GeV]}&\multicolumn{2}{c@{}}{\makecell{Beam Energy {}[GeV]}} &
      \makecell{$p_T$ range}
      & \multicolumn{2}{c}{Model I} & \multicolumn{2}{c}{Model II} 
      \\
      \cline{3-4}\cline{6-9} \cline{10-11}
      &&p&e&
      [GeV]
      &$\sigma\,[\mathrm{fb}]$&$N$&$\sigma\,[\mathrm{fb}]$&$N$\\
      \hline
      \multirow[c]{4}{*}{EIC} & 44.7  & 100   & 5    & 6-20 & $0.022$	&	$2.2$	&	$0.0031$	&	$0.31$	\\
                              & 63.2  & 100   & 10   & 6-20 &  $0.069$	&	$6.9$	&	$0.0098$	&	$0.98$	\\
                              & 104.9 & 275   & 10   & 6-20 &  $0.25$	&	$25.$	&	$0.035$	&	$3.5$	\\
                              & 140.7 & 275   & 18   & 6-20 &  $0.45$	&	$45.$	&	$0.064$	&	$6.4$	\\
                              \hline
      HERA                    & 319   & 920   & 27.5 & 6-20 &  $1.5$	&	$0.72$	&	$0.22$	&	$0.10$	\\
      \hline
      EicC                    & 20    & 19.08 & 5    & 6-9  &  $0.000015$	&	$0.00076$	&	$2.1\times 10^{-6}$	&	$0.00011$	\\
      \hline \hline
    \end{tabular}
    \caption{The $p_T$-integrated cross section for $T_{4c}$ inclusive production at the EIC. The integrated luminosity of the \texttt{EIC} is assumed to be $100\ \mathrm{fb}^{-1}$ for one year of data taking, as opposed to $50.5\ \mathrm{fb}^{-1}$ for the \texttt{EicC}. The integrated luminosity of \texttt{HERA} is $468\ \mathrm{pb}^{-1}$. The estimated event yields for the \texttt{EIC} and \texttt{EicC} only account for numbers per year. }
    \label{tab:total:cross:section}
\end{table}
\endgroup

Finally in Table~\ref{tab:total:cross:section} we enumerate the integrated cross sections and event yields of the vector $T_{4c}$ at the \texttt{EIC}, \texttt{HERA}, and \texttt{EicC}. 
The integrated luminosity of \texttt{HERA} is $468\ \mathrm{pb}^{-1}$, while that of the \texttt{EIC} is assumed to be $100\ \mathrm{fb}^{-1}$ for one year of data taking, 
as opposed to $50.5\ \mathrm{fb}^{-1}$ for the \texttt{EicC}. 
The $p_T$ is integrated over the range $6\ \mathrm{GeV}\leq p_T\leq 20\ \mathrm{GeV}$ for the \texttt{EIC} and \texttt{HERA}. 
Due to the small beam energy, we set the $p_T$ range to be $6\ \mathrm{GeV}\leq p_T\leq 9\ \mathrm{GeV}$ for the \texttt{EicC}. 
From Table~\ref{tab:total:cross:section} we see that the event yields per year appear to be considerable at the the \texttt{EIC},  which implies that the observation potential of the
vector $T_{4c}$ at future \texttt{EIC} experiment may be promising. 
However,  the observation prospects at both \texttt{HERA} and the \texttt{EicC} look gloomy, which may be largely attributed to the low luminosity of \texttt{HERA}
and the small beam energy of the \texttt{EicC}.

\section{Summary \label{sec:summary}}

In this paper, within the NRQCD factorization framework, we predict the $p_T$ distributions for the inclusive photoproduction of the fully charmed tetraquark 
at electron-proton collisions with different beam energies, at lowest $\alpha_s$ and $v$. Due to the $C$-parity conservation, only the $1^{+-}$ $T_{4c}$ can be produced through the photon-gluon
fusion process. With the LDME estimated through phenomenological potential models, we also predict the integrated production rates and event yields of the vector $T_{4c}$ 
at the \texttt{EIC}, \texttt{HERA}, and \texttt{EicC}. Our study suggests that the \texttt{EIC} is the most promising $ep$ collider for detecting the vector $T_{4c}$ events.

\begin{acknowledgments}
The work of F.~F. is supported by the National Natural Science Foundation of China under Grants No.~12275353 and No.~11875318.
The work of Y.-S.~H. is supported by DOE Grants No. DE-FG02-91ER40684 and No. DE-AC02-06CH11357.
The work of Y.~J. and J.-Y.~Z. is supported in part by NNSFC Grants No.~11925506 and No.~12070131001 (CRC110 by DFG and NSFC).
The work of  W.-L. S. is supported in part by the National Natural Science Foundation of China under Grants No. 12375079 and No. 11975187, and the Natural Science Foundation of ChongQing under Grant No. CSTB2023NSCQ-MSX0132.
The work of D.-S. Y. is supported by the National Natural Science Foundation of China under Grants No. 12235008, and by the Ministry of Science and Technology of China under Grants No. 2022YFA1601900. 
The work of J.-Y.~Z. is also supported in part by U.S. Department of Energy (DOE) Contract No.~DE-AC05-06OR23177, under which Jefferson Science Associates, LLC, operates Jefferson Lab.
\end{acknowledgments}


\end{document}